\documentclass[twocolumn,a4paper,preprintnumbers,amsmath,amssymb,nofootinbib,floatfix]{revtex4}

\usepackage{color}
\usepackage{graphicx}
\usepackage{dcolumn}
\usepackage{bm}
\usepackage{latexsym}
\usepackage{epsfig}
\usepackage{rotating}
\usepackage{bm}

\newcommand{\be}{\begin{equation}}
\newcommand{\ee}{\end{equation}}
\newcommand{\beqq}{\setlength\arraycolsep{2pt}\begin{eqnarray}}
\newcommand{\eeqq}{\vspace{0cm} \end{eqnarray}}
\newcommand{\bea}{\begin{eqnarray}}
\newcommand{\eea}{\end{eqnarray}}

\begin{document}

\title{Evolution of the universe driven by a mass dimension one fermion field}

\author{S. H. Pereira$^{1}$, R. F. L. Holanda$^{2,3}$ and A. Pinho S. Souza$^{1}$}

\affiliation{$^{1}$Universidade Estadual Paulista (Unesp)\\Faculdade de Engenharia, Guaratinguet\'a \\ Departamento de F\'isica e Qu\'imica\\ Av. Dr. Ariberto Pereira da Cunha 333\\
12516-410 -- Guaratinguet\'a, SP, Brazil\\
$^{2}$Universidade Federal de Sergipe, Departamento de F\'{\i}sica,  49100-000 -- Aracaju, SE, Brazil,\\
$^{3}$Universidade Federal de Campina Grande, Departamento de F\'{\i}sica,  58429-900 -- Campina Grande, PB, Brazil
}




\begin{abstract}

This paper study the evolution of the universe filled with a neutral mass dimension one fermionic field, sometimes called Elko. The numerical analysis of the coupled system of equations furnish a scale factor growth and energy density evolution that correctly reproduces the inflationary phase of the universe. After that, supposing a mechanism of energy transference  to ordinary matter, the initial conditions generated after inflation drives the radiation dominated phase and also the subsequent dark matter evolution, since the Elko field is a good dark matter candidate. The energy density of the field at the end of inflation, at the end of radiation phase and for present time are in agreement to the standard model estimates. The analysis was performed with a potential containing  a quadratic mass term plus a quartic self-interaction term, which follows naturally from the theory of mass dimension one fermions. It is interesting to notice that inflation occurs when the field makes a kind of transition around the Planck mass scale. The number of e-foldings during inflation was found to be strongly dependent on the initial conditions of the Elko field, as occurs in chaotic inflationary models. An upper mass limit for Elko field has been obtained as $m<10^9$GeV. A possible interpretation of both inflationary phase and recent cosmic acceleration as a consequence of a kind of Pauli exclusion principle is presented at the end. 

\end{abstract}

\maketitle

\section{Introduction}

The search for a model that correctly describes the whole evolution of the universe is an old problem in cosmology. In the current model, the universe starts in a very hot and dense phase known as big bang\footnote{See \cite{kolb} or \cite{weinberg} for a brief resume of the thermal history of the universe.} driven by quantum effects based on models as supersymmetry, supergravity, extra dimensions, superstrings, among others. The quantum effects are dominant while the energy density is greater than Planck energy density $m_{pl}^4$, characterized by the Planck mass $m_{pl}\simeq 1.22\times 10^{19}$GeV. The age of the universe is about $10^{-43}$s at this stage. From $10^{-43}$s to about $10^{-35}$s the universe is still expanding and cooling in the so called pre-inflationary phase, its temperature is about $10^{27}$K and the energy density is about $10^{58}$GeV$^4\sim$ $10^{75}$g/cm$^3$. From $10^{-35}$s to $10^{-32}$s the universe undergoes the so-called {cosmic} inflation, where the scale factor $a(t)$ growth for about $10^{43}$ orders of magnitude. Such phase is necessary in order to solve some problems as the flatness problem, the monopole and relics problems, the horizon problem and homogeneity \cite{bookliddle,linde1}. After such very abrupt expansion, nearly exponential, the energy density is about $10^{51}$GeV$^4\sim 10^{68}$g/cm$^3$ and the universe {goes through} a reheating phase, where the so called inflaton field transfer energy to the ordinary matter, which starts to dominate. After $10^{-32}$s the universe goes into radiation dominated phase up to $10^{11}$s, and several process occur, as the end of electroweak unification, the quark-hadrons transition and nucleosynthesis of light elements. At the end of radiation dominated phase, radiation and matter has the same energy density, of about $10^{-34}$GeV$^4\sim 10^{-17}$g/cm$^3$. Then the universe enters a matter dominated {long epoch}, {in which} occurs the formation of atoms and first structures, as galaxies and cluster of galaxies.  The scale factor growth about $10^{65} - 10^{75}$ orders of magnitude up today, with an age of about $10^{18}$s and the energy density decreases to about $10^{-46}$GeV$^4\sim 10^{-29}$g/cm$^3$. Finally, very recently the universe starts a new accelerating phase, dominated by a cosmological constant term or a dark energy fluid. This is a very brief history of the universe.

An unified model that could describe all the phases of evolution of the universe is a difficult task. The several orders of magnitude involved from the inflation to recent cosmic acceleration forced the researches to divide the evolution of the universe into different parts, each one characterized by different kind of particles that dominate at different stages. These ingredients form the so called standard model of cosmology. The inflationary phase of the universe can be constructed with a single scalar field which drives the inflation while the scalar field rolls down to the bottom of its potential. Several potentials that satisfies the fine tunings of the inflationary phase have been studied in last decades (see \cite{bookliddle,linde1} for a review and \cite{Planck2013} for observational constraints on several potentials.). After inflation, the scalar field ends in a rapid oscillation around the minimum of its potential and its energy is transferred to the baryonic particles in the next phase of evolution, in a process known as reheating.  The scalar field does not act anymore and the universe evolves dominated by radiation and subsequently by matter. During the matter dominated phase there is the necessity {of adding} a new kind of non-baryonic matter to the model in order to correctly explain structure formation, thus the standard model needs addition of dark matter, about 25\% of the total content of the universe. Radiation and baryons corresponds just to about 5\% of the total content. After evolving dominated by dark matter, observations show that a recent {accelerated} phase of expansion needs a new ingredient, the so called cosmological constant or even a new kind of energy, called dark energy, acting as a vacuum energy density and representing about 70\% of the total energetic content.

As previously stated, a model describing all phases of the evolution of the universe is a challenge for the present cosmology. In particular, the constraints in the energy density and the `size' of the universe after each phase is strongly dependent on the inflationary model adopted. The exact initial conditions {leading a cosmic} inflation and its connection to radiation and matter evolution depends on the dynamic of the inflaton field. Recently, a new class of fermions with mass dimension one named Elko \cite{AHL1,AHL2,ahl2011a,AHL4} was proposed and established in solid quantum bases very recently \cite{AHL4,WT} (see also \cite{RJ}). Such new fermionic field is a natural candidate to dark matter in the universe once it is constructed by means half-integer spinors that are eigenstate of the charge conjugation operator, being neutral and weakly coupled to the electromagnetic sector of the standard model of particles. Several cosmological applications of such fermionic mass dimension one field have been recently presented in the literature \cite{FABBRI,BOE4,BOE6,GREDAT,BASAK,sadja,kouwn,saj,js,asf,sajf,st}. 
The great advantage to use the Elko field as the inflaton field is that it does not need to disappear after acting, since it happens to be responsible for the dark matter in the present day. Also, being a fermionic field, its quantum properties could be evoked in order to better understand both the inflationary phase as recent cosmic acceleration, as a kind of Pauli exclusion principle mechanism.
 
In the present paper we consider the fermionic Elko field as an alternative to drive the inflationary phase of the universe. If we suppose that it transfers only part of its energy to radiation after inflation\footnote{The mechanism of such transfer is not known yet, but the possible coupling of Elko field to Higgs field has been studied recently \cite{AHL2,dias1,alves1}, thus the Higgs field could intermediate such energy transfer to ordinary matter.}, the evolution of universe up to present time can be correctly described, namely the end of inflation serves as initial conditions to radiation phase and the end of radiation as initial conditions to matter dominated phase. We present the numerical results concerning the Elko field equations obtained recently in \cite{sajf} and here driving the inflationary phase in the presence of a potential with a quadratic mass term and a quartic self interaction term, similar to which has been done recently in \cite{st}. It is shown that inflation occurs when the Elko field makes a transition characterized by a value of about one Planck mass. The Planck scale represents a limit from pre-inflationary phase to post-inflationary one. The evolution for the scale factor and the energy density during and after {a cosmic} inflation are in good agreement to the values of the standard model for three different epochs, namely the end of inflation, the transition from radiation to matter dominance and the present day value for such parameters. It is important to stress that the time scale in seconds used here allows to estimate the real growth of the scale factor and the decrease of the energy density up today. As quoted above, a similar treatment has been done recently in \cite{st} concerning the numerical analysis at the inflationary phase, however using a phenomenological symmetry broken potential, which is not predicted by mass dimension one fermions theory \cite{AHL2}. Also, in \cite{st} the evolution after inflation was just extrapolated, not covering the radiation and matter evolution from its known evolution equations, as done here.

\begin{figure*}[ht]
\centering
\includegraphics[width=0.9\textwidth]{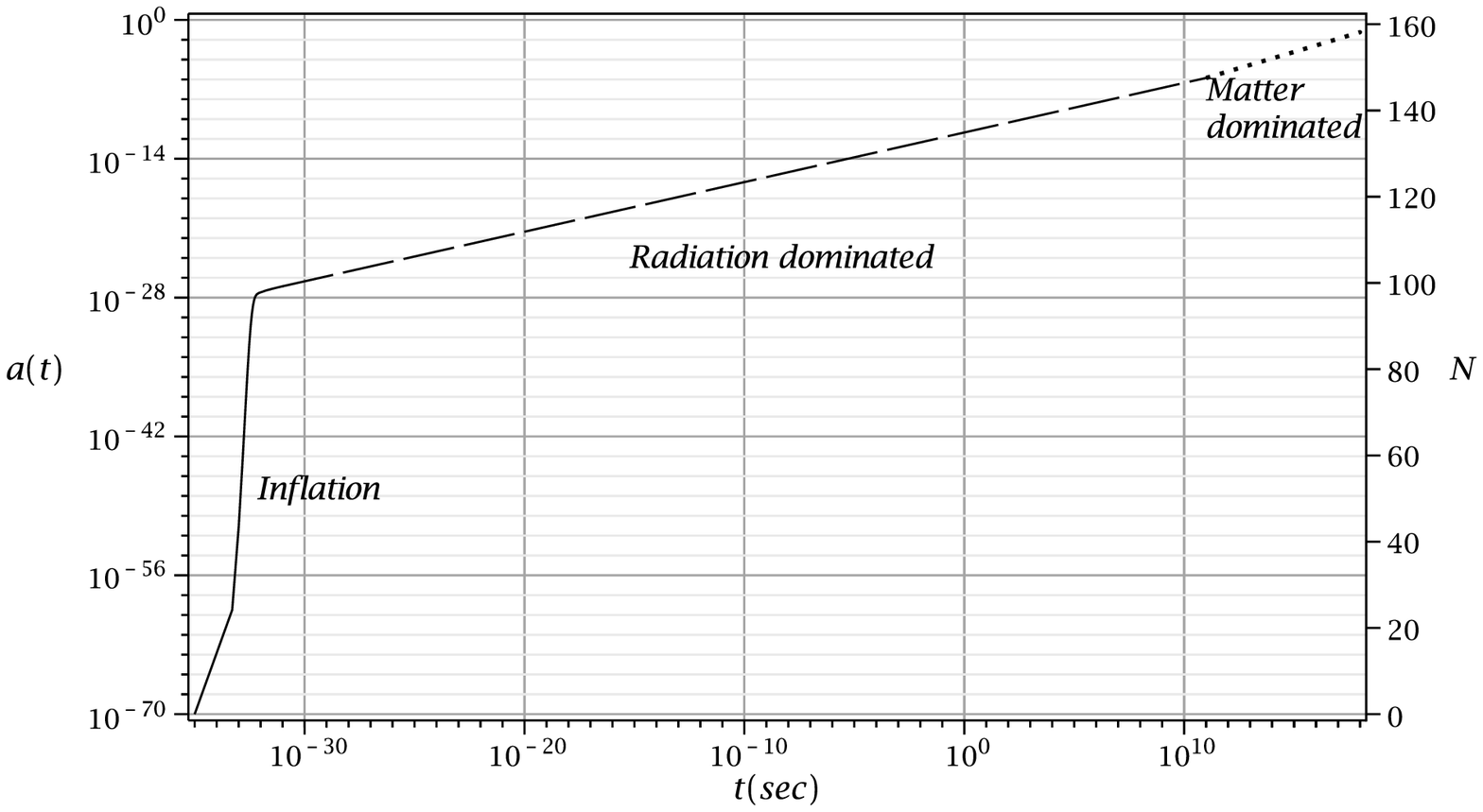}
\caption{Numerical result for the evolution of the scale factor $a(t)$ with time. Inflation occurs from $10^{-35}$s to $10^{-32}$s (solid line) with the scale factor increasing from $10^{-70}$ up to about $10^{-28}$. After that the scale factor grows smoothly dominated by radiation (dashed line), with $a(t)\sim t^{1/2}$ up to about $t\simeq 10^{11}$s. From  $10^{11}$s to $10^{18}$s the evolution is of matter type (dotted line), with $a(t)\sim t^{2/3}$. The growth of the scale factor up today, $a(t)\sim 1$, corresponds to an increase of about $10^{70}$. The vertical axis on the right shows the number $N$ of e-foldings.}
\end{figure*}

\begin{figure*}[ht]
\centering
\includegraphics[width=0.9\textwidth]{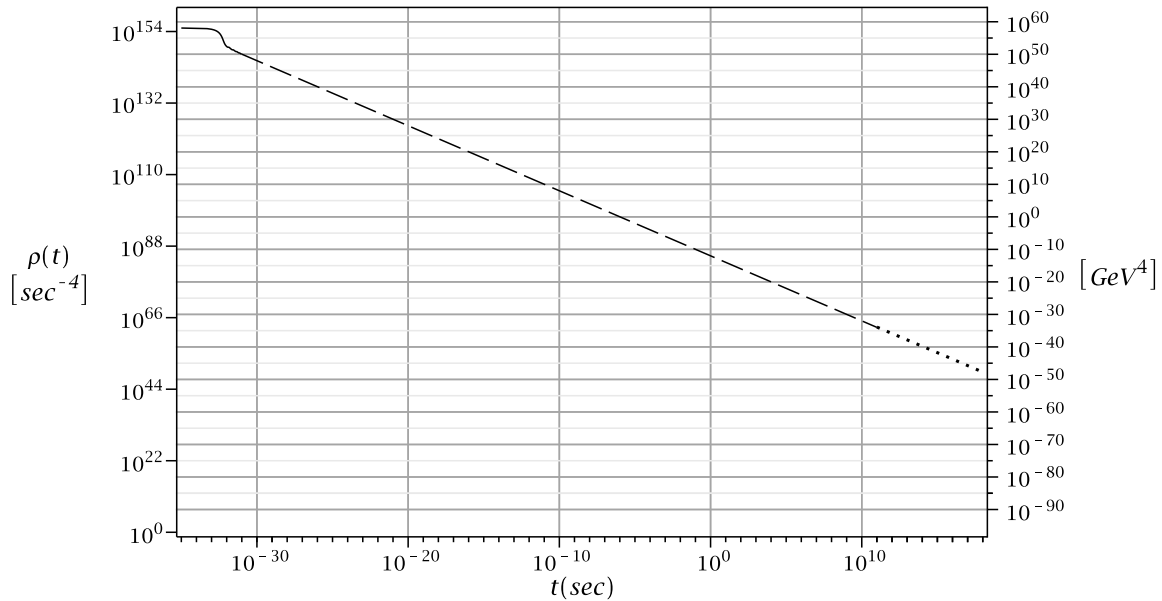}
\caption{Numerical result for the evolution of the energy density (\ref{rhoElko}) of the Elko field with time. The units of vertical scale on the left is [s$^{-4}$] and on the right is GeV$^4$. During the inflationary phase the energy density is nearly constant, about $10^{155}$s$^{-4}\simeq 10^{58}$GeV$^4\simeq 10^{75}$g/cm$^3$. The inflation finish at $10^{-32}$s with an energy density of about $10^{149}$s$^{-4}\simeq 10^{52}$GeV$^4\simeq 10^{69}$g/cm$^3$. Up to $t\simeq 10^{-30}$s (solid line) the figure was obtained numerically using the solution for $\phi(t)$ into (\ref{rhoElko}). After $t\simeq 10^{-30}$s (dashed line) the evolution corresponds to the radiation energy density of the form $\rho_{r}=\rho_{r0}a(t)^{-4}$, with $\rho_{r0}=4\times 10^{39}$s$^{-4}$, showing the reduction of the radiation energy density up to $10^{11}$s, where matter and radiation are in equilibrium, with an energy density of about $10^{-35}$GeV$^4\simeq 10^{-18}$g/cm$^3$. After that the evolution follows the matter energy density (dotted line) as  $\rho_{m}=\rho_{m0}a(t)^{-3}$, with $\rho_{m0}=3.5\times 10^{35}$s$^{-4}$, up  today, where the energy density is about $10^{51}$s$^{-4}\simeq 10^{-46}$GeV$^4\simeq 10^{-29}$g/cm$^3$.}
\end{figure*}

\begin{figure*}[ht]
\centering
\includegraphics[width=0.325\textwidth]{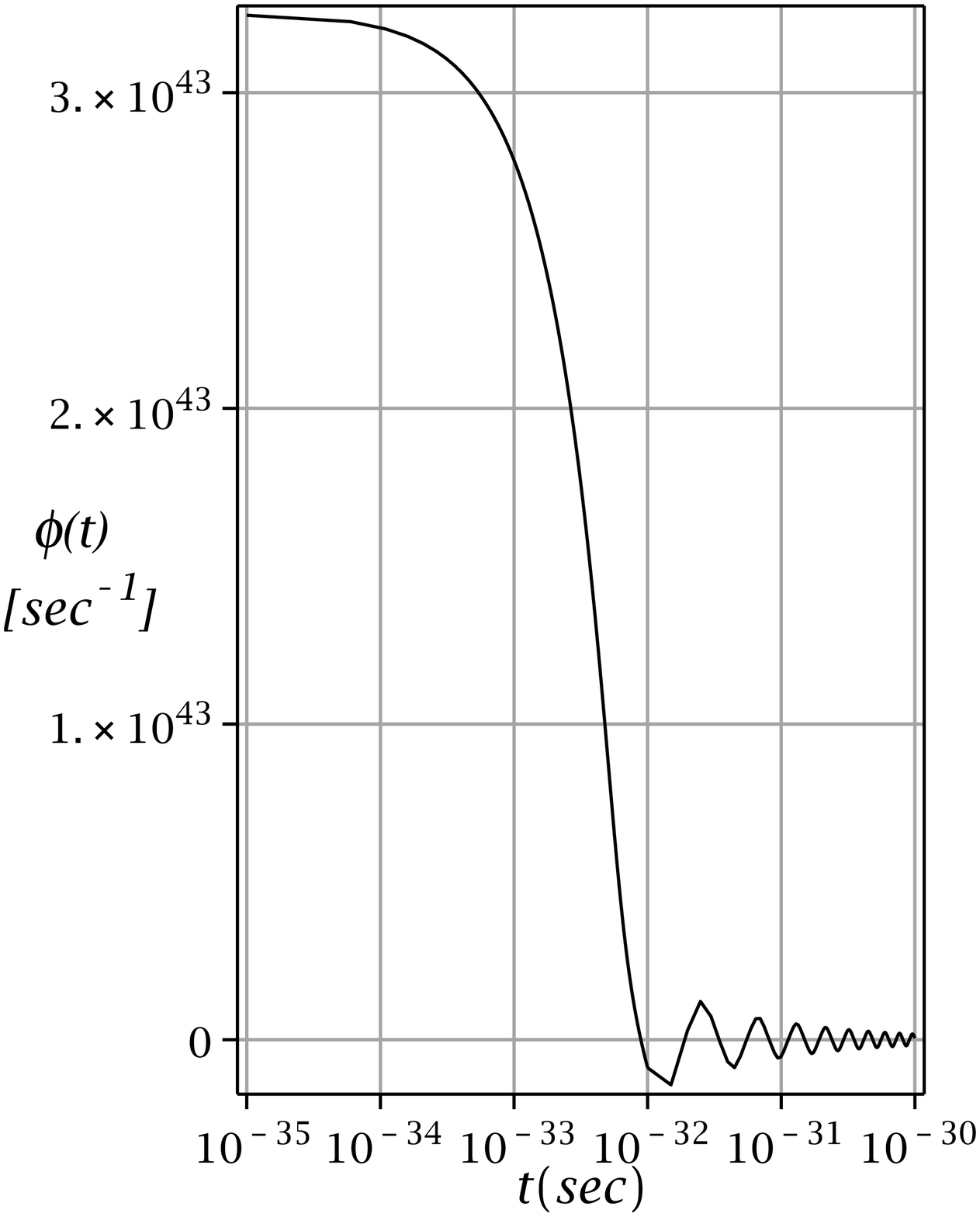}
\hspace{0.3cm}
\includegraphics[width=0.28\textwidth]{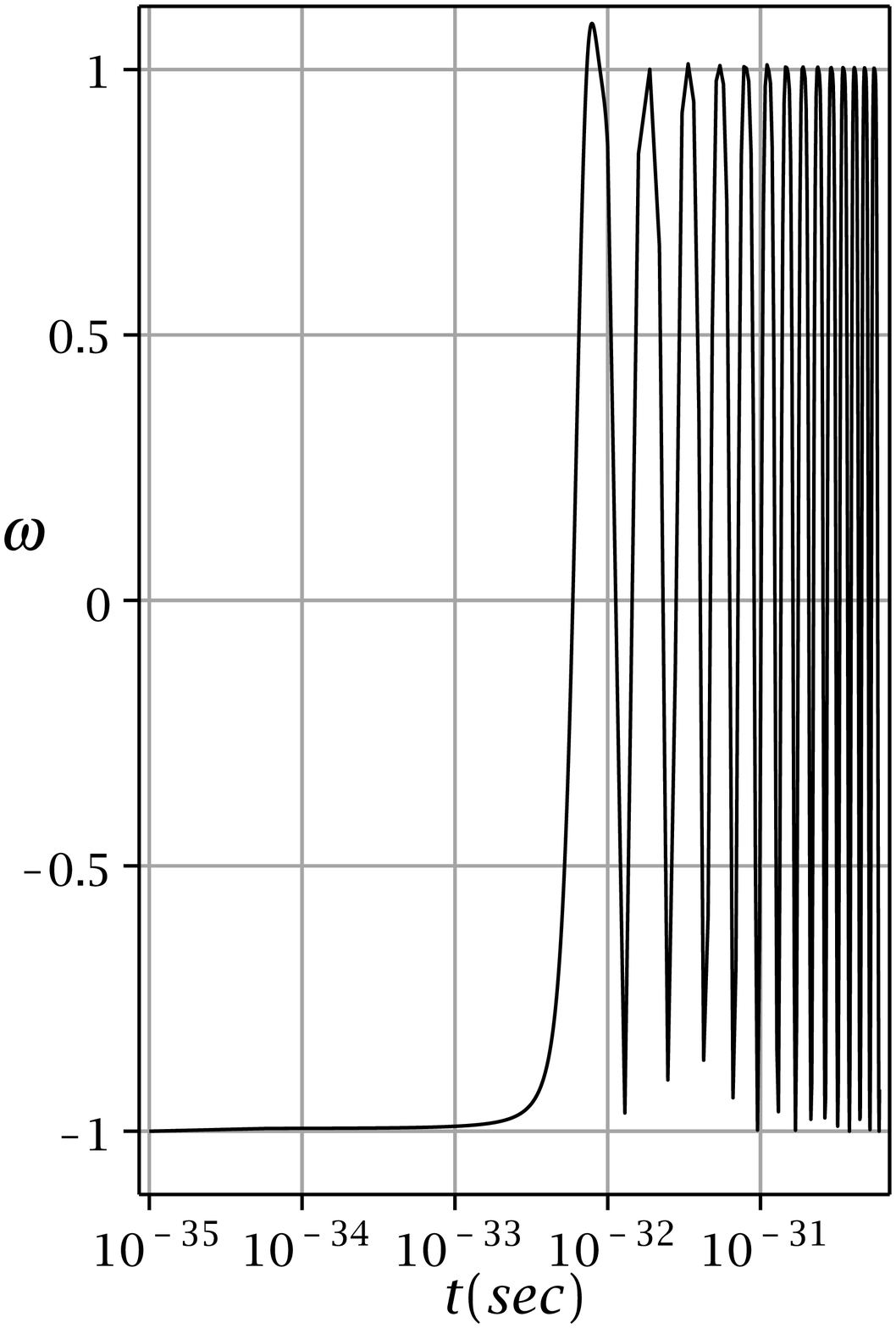}
\hspace{0.3cm}
\includegraphics[width=0.32\textwidth]{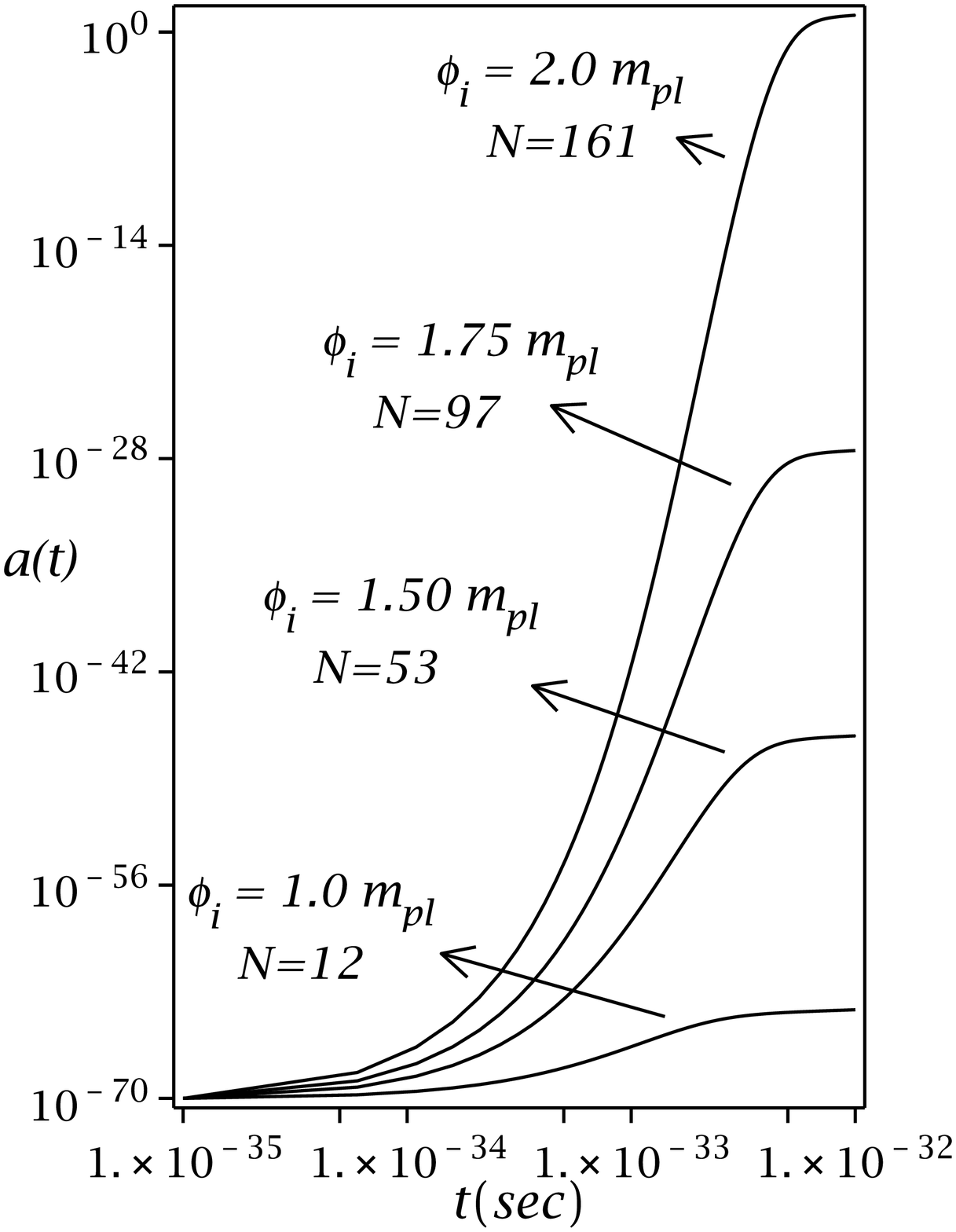}\\
\hspace{0.7cm}(a)\hspace{5.5cm}(b)\hspace{5.4cm}(c)
\caption{(a) - Evolution of the Elko field $\phi(t)$ to its minimal value showing its oscillation with time while rolling down to the bottom of the potential. (b) - Evolution of the equation of state parameter $\omega=p/\rho$ with time, showing its oscillation around 0 after starts in $\omega=-1$. (c) - Detailed view of the inflationary phase and its strong dependence on the initial value $\phi_i$. The number $N$ of e-foldings are also presented to each case. }
\end{figure*}

\section{Dynamic equations for Elko field}

In a flat space-time there are four Elko spinors satisfying invariance by the charge conjugation operator $C$, they are labeled as $\lambda^{S/A}_\beta$ where $S$ stands for Self-conjugate and $A$ for Anti-self-conjugate. The index $\beta$ stands for two possible helicities. They satisfies $C\lambda^{S/A}_\beta = \pm \lambda^{S/A}_\beta$ \cite{AHL1,AHL2,ahl2011a,AHL4,RJ,WT} and are normalized satisfying the relation $\stackrel{\neg}{\lambda}_{\beta}({\bf k})^{S/A}\lambda_{\beta^{'}}({\bf k})^{S/A} = \pm 2m\delta_{\beta\beta^{'}}$, where $\lambda_{\beta^{'}}({\bf k})^{S/A}$ and $\stackrel{\neg}{\lambda}_{\beta}({\bf k})^{S/A}$ are the usual spinor and its dual, respectively. The dual have been redefined recently \cite{AHL4,RJ,WT} in order to maintain locality and Lorentz covariance. The corresponding quantum fields can be redefined in order to satisfies a normalization relation as $\stackrel{\neg}{\lambda}\lambda = \pm 1$, where $\lambda$ stands here for any of the four spinors, with normalization $+1$ for the two self-conjugate and $-1$ for the two anti-self-conjugate. 

In this paper it was used just one fermionic Elko field satisfying a positive normalization. Also, in order to use the Elko field in a curved background, it was factored out the time dependence of the Elko field as \cite{BOE4,BOE6,GREDAT,
BASAK,sadja,kouwn,sajf,saj,js,asf,st} $\Lambda=\phi(t) \lambda$. The action for the model reads \cite{kouwn,sajf}:
\begin{equation}
S = \int d^4 x \sqrt{-g} \left[ -\frac{\tilde{R}}{2\kappa^2}+{1\over 2}g^{\mu\nu}\tilde{\nabla}_\mu \stackrel{\neg}{\Lambda}\tilde{\nabla}_\nu \Lambda -V(\stackrel{\neg}{\Lambda}\Lambda) \right] \,,
\label{actionE}
\end{equation}
where $\kappa^2\equiv8\pi G$ with $c=1$. The tilde denotes the presence of torsion terms into the Ricci scalar $\tilde{R}$ and covariant derivatives, namely, $\tilde{\nabla}_\mu \Lambda\equiv \partial_\mu\Lambda - \Gamma_\mu \Lambda$ and $\tilde{\nabla}_\mu \stackrel{\neg}{\Lambda} \equiv \partial_\mu\stackrel{\neg}{\Lambda} + \stackrel{\neg}{\Lambda}\Gamma_\mu$,
where $\Gamma_\mu$ is the connection associated to spinor fields, containing the spin connections. For the potential it was used a quadratic mass term and a quartic self-interacting term, namely:
\begin{equation}
V= \frac{1}{2}m^2 \stackrel{\neg}{\Lambda}\Lambda + {\alpha\over 4}(\stackrel{\neg}{\Lambda}\Lambda)^2 =  \frac{1}{2}m^2\phi^2 + {\alpha\over 4}\phi^4\,,\label{potential}
\end{equation}
where $m$ is the physical mass of the field and $\alpha$ a dimensionless coupling. Such kind of potential follows naturally from the theory of mass dimension one fermions \cite{AHL2}. 

In a flat Friedmann-Robertson-Walker (FRW) metric $ds^2=dt^2-a(t)^2[dx^2 + dy^2 + dz^2]$, the two Friedmann equations and the dynamic field equation for the time component $\phi(t)$ of the Elko field\footnote{We will refer to $\phi(t)$ as Elko field from now on, but the complete spinor structure is present in $\Lambda$.} plus a ordinary matter $i$ can be obtained \cite{kouwn,sajf}:
\begin{equation}
H^2={\kappa^2\over 3}\bigg(1+{\kappa^2\phi^2\over 8} \bigg)\bigg[{\dot{\phi}^2\over 2}+V(\phi)+\rho_i\bigg]\,,\label{H2}
\end{equation}
\begin{equation}
\dot{H}=-{\kappa^2\over 2}\bigg(1+{\kappa^2\phi^2\over 8} \bigg)\bigg[\dot{\phi}^2-{1\over 2}{H\phi\dot{\phi} \over (1+\kappa^2\phi^2/8)^2} +\rho_i+p_i\bigg]\,,\label{Hdot}
\end{equation}
\begin{equation}
\ddot{\phi}+3H\dot{\phi}+{dV(\phi)\over d\phi}-{3\over 4}{H^2\phi\over (1+\kappa^2\phi^2/8)^2}=0\,,\label{phiElko}
\end{equation}
where $H={\dot{a}\over a}$ is the Hubble parameter and a dot stands for time derivative. The energy density $\rho_i$ and the pressure $p_i$ for the ordinary matter $i$ (radiation or baryonic matter) where also included in the right side of (\ref{H2}) and (\ref{Hdot}). The equations for energy density and pressure for Elko field are \cite{sajf}:
\begin{equation}
\rho={\dot{\phi}^2\over 2}+V(\phi)+{3\over 8}{H^2\phi^2\over (1+\kappa^2\phi^2/8)}\,,\label{rhoElko}
\end{equation}
\begin{eqnarray}
p&=&{\dot{\phi}^2\over 2}-V(\phi)-{3\over 8}{H^2\phi^2\over (1+\kappa^2\phi^2/8)}- {1\over 4 }{\dot{H}\phi^2\over (1+\kappa^2\phi^2/8)}\nonumber\\ &&-{1\over 2}{H\phi\dot{\phi}\over (1+\kappa^2\phi^2/8)^2}\,.\label{pphiElko}
\end{eqnarray}

Such equations generalizes that one for a standard scalar field. The second terms inside curl bracket of (\ref{H2})-(\ref{Hdot}), the second term inside square bracket of (\ref{Hdot}), the last terms of (\ref{phiElko}) and (\ref{rhoElko}) and last three terms of (\ref{pphiElko}) are not present in the standard scalar field equations. This shows that the Elko dynamic field equations are much richer than the scalar field ones. Such additional terms comes from the fact that the fermionic Elko field must be coupled to torsion in an Einstein-Cartan framework and also due to spin connections terms. In particular, since that $\kappa^2 = 8\pi G =8\pi/m_{pl}^2$, in the limit $\phi\ll m_{pl}$ and $H\phi \ll \dot{\phi}$ these additional terms can be discarded and the equations are exactly like that ones for the standard scalar field. Thus the limit $\phi \sim m_{pl}$ represents a kind of transition of the Elko field from a high energy regime dominated by torsion and spin connection terms to a low energy regime free of torsion. Surprisingly, it has been found numerically that such transition from a high energy regime to a low energy one is the responsible for the inflationary phase of the universe. Notice also that in the limit $\phi\ll m_{pl}$ we have the torsion term $\kappa^2\phi^2/8\ll 1$, which simplifies the system of equations, being very similar to the standard scalar field, but it is important during inflationary phase.

Finally, given a potential $V(\phi)$ the above set of equations can be integrated (at least numerically) in order to study the evolution of $a(t)$ and $\phi(t)$. In which follows it will set the initial conditions and values of the parameters in order to make a numerical analysis of the system.

\section{Numerical results}

In order to perform a numerical analysis\footnote{It was used the Maple 15 Software, whose numerical solutions are found by a Fehlberg fourth-fifth order Runge-Kutta method with degree four of interpolation.} for the evolution of the scale factor $a(t)$ and $\phi(t)$ it was chosen to work with the coupled equations (\ref{H2}) and (\ref{phiElko}). It is also necessary to fix some parameters of the potential (\ref{potential}) and initial conditions for $\phi$. Since (\ref{H2}) is a first order differential equation in $a(t)$ it needs just one initial condition at $t_i$, and it was chosen $a(t_i)=10^{-70}$. With such choice we expect the present value to be $a=1$. Equation (\ref{phiElko}) is a second order differential equation and needs two initial conditions. In order to test the hypothesis that inflation occurs due to a transition of the Elko field from a high energy scale $\phi > m_{pl}$ to a low energy regime $\phi < m_{pl}$, it was chosen $\phi(t_i)\equiv \phi_i = 1.75m_{pl}=3.25\times 10^{43}$s$^{-1}$, similar to chaotic inflationary model by Linde \cite{lindePLB}. The inflation must occur when $\phi(t)$ decays from $\phi_i$ to zero, to the bottom of the potential. It has been also considered that before inflation occurs the universe was filled with a nearly homogeneous and isotropic gas of Elko particles, nearly at rest, so it was chosen $\dot{\phi}(t_i)\equiv \dot{\phi}_i=0$. Thus the unique energy present at the beginning is its potential energy $V(\phi_i)$. During inflation we do not consider the presence of ordinary matter $\rho_i$ nor $p_i$.

After some numerical analysis it was found that the beginning of the evolution, namely the phase that includes the inflation, can be driving just by the self-interacting quartic term of the potential if $m^2\ll {1\over 2} \alpha \phi_i^2$, which is obvious since that it is proportional to $\phi^4$ while the mass term is proportional to $\phi^2$ and the initial condition for $\phi$ is nearly greater than the Planck mass. By choosing $\alpha = 4\times 10^{-20}$ the condition for the mass is $m\ll 2.5\times 10^{-10}m_{pl} \simeq 3\times 10^9$GeV. It was chosen $m=1.0$GeV so that it does not contribute to the evolution at the beginning. When the field decays below about $10^{-10}m_{pl} \simeq 10^{33}$s$^{-1}$ the quadratic mass term starts to dominate. Numerically this occurs at about $10^{-5}$s, being important already in radiation dominated epoch.  Having stated the initial conditions and the value of the parameters, we see that the condition $V(\phi_i)\ll m_{pl}^4$ is satisfied in order to warranty that the total energy of the field is below the Planck scale, so that quantum effects are not present. Such condition is satisfied due to tiny value of $\alpha$. Now it is possible to make a numerical analysis of the system of equations (\ref{H2}) and (\ref{phiElko}). In order to have the time scale in seconds, it was used $\kappa=\sqrt{8\pi}/m_{pl}= 4.1\times 10^{-19}$GeV$^{-1} = 2.7\times 10^{-43}$s. The initial time $t_i$ was chosen as $t_i=10^{-35}$s. It is expected that inflation occurs up to about $10^{-32}$s and after that the universe expand in a nearly power law in time.

Figures 1 and 2 present the main results of the paper. The black solid lines correspond to numerical analysis just for Elko field, namely with $\rho_i=0$ into \eqref{H2}. In Figure 1 the evolution of the scale factor $a(t)$ with time (in seconds) for the above initial conditions and parameters of the potential shows that inflation in fact occurs up to about $10^{-32}$s and the scale factor growth for about $10^{42}$ orders of magnitude, which corresponds to about 97 e-foldings, calculated as $N=\ln a(t)/a(t_i)$ and indicated on the vertical axis on the right. After that, by supposing the Elko field transfer part of its energy to ordinary matter, the universe evolves as $a(t)\sim t^{1/2}$ in a radiation dominated phase, which is indicated by the strait dashed line in the logarithm scale. After that, once radiation finish its dominance, the rest of Elko field that was also evolving together radiation act again, as a dark matter with null pressure and $a(t)\sim t^{2/3}$. The evolution follows up today with the scale factor of about 1, our normalization for present day, in good agreement to the previous estimate from \cite{kolb} of an increase of about $10^{74}$ for the standard model.

Figure 2 shows the numerical analysis for the evolution of the energy density $\rho(t)$ from (\ref{rhoElko}), showed in solid line. It can be seen that during the inflationary phase (from $t\sim 10^{-35}$s to $10^{-32}$s) the energy density is nearly constant, about $10^{155}$s$^{-4}\simeq 10^{58}$GeV$^4\simeq 10^{75}$g/cm$^3$. The inflation finishes with an energy density of about $10^{149}$s$^{-4}\simeq 10^{52}$GeV$^4\simeq 10^{69}$g/cm$^3$, in good agreement with standard model \cite{kolb} of about $10^{69}$g/cm$^3$. After $t\simeq 10^{-30}$s (dashed line) the evolution corresponds to the radiation energy density of the form $\rho_{r}=\rho_{r0}a(t)^{-4}$, showing the reduction of the radiation energy density up to $10^{11}$s, where matter and radiation are in thermal  equilibrium.  At this stage the energy density if of about $10^{-35}$GeV$^4\simeq 10^{-18}$g/cm$^3$. After that the evolution follows the matter energy density (dotted line) as  $\rho_{m}=\rho_{m0}a(t)^{-3}$ up  today, where the energy density is about $10^{50}$s$^{-4}\simeq 10^{-47}$GeV$^4\simeq 10^{-29}$g/cm$^3$, exactly the expected value for the critical energy density today. The energy density at all transitions phase are in good agreement to the standard model.

Figure 3 (a) shows the numerical results for the decaying of the Elko field $\phi(t)$ from its initial value $\phi_i=1.75m_{pl}$ with time. Before reach the bottom of the potential the field oscillates for a long time around its minimal value. Figure 3 (b) shows the numerical analysis of the equation of state parameter $\omega \equiv p(\phi)/\rho(\phi)$ obtained with (\ref{rhoElko}) and (\ref{pphiElko}), which indicates that $\omega$ starts from $-1$ during inflation and then begins to oscillate around $\omega=0$ for the rest of its evolution. This confirms that the Elko field satisfies a dust equation of state type, exactly as desired for a dark matter fluid. Thus the Elko field act as a dark matter after radiation dominated era, up to the rest of the evolution of the universe. Also, having nearly null kinetic energy the Elko particles could be attracted to other local potentials, initiating the growth of small structures, as  dark matter halos. Remember that it is one the main characteristic of the Elko field, a candidate to dark matter particle. Figure 3 (c) shows in detail what happens during the inflationary phase for different initial conditions of $\phi$, indicating a strong dependence with the initial value $\phi_i$ of the field. The other parameters are the same of previous analysis. It is possible to see clearly that the number of e-foldings of the inflation is strongly dependent on the initial condition for the field. This is a kind of fine tuning for the model and shows that inflation occurs when the field decreases below the Planck mass scale.

\section{Concluding remarks}

The numerical analysis of the coupled system of equations concerning a homogeneous and isotropic distribution of mass dimension one fermionic Elko field filling the whole universe was performed, considering the Elko field as an alternative to standard scalar field inflationary model. The potential under which the Elko field slows down is a quadratic mass term plus a quartic self-interaction. It was found that the evolution for the scale factor reproduces the expected exponential growth during the inflationary phase up to about $10^{-32}$s, with the desired number of e-foldings. The energy density after inflation is also in agreement with standard model. After inflation ends, by supposing a partial energy transference from Elko field to radiation, the values for the scale factor and energy density serve as initial conditions for the radiation phase. Surprisingly the evolution of the type $t^{1/2}$ for the scale factor and $a(t)^{-4}$ for the energy density leads the universe to the expected value for the energy density at $10^{11}$s, where matter and radiation are in equilibrium. Being the Elko field of matter type, with null pressure, it dominates again, given rise to the dark matter phase in the universe, evolving as $t^{2/3}$ for the scale factor and $a(t)^{-3}$ for the energy density. Again, the present day value for the energy density is obtained in agreement with the standard model, and the total growth for the scale factor is about $10^{70}$, also in agreement to standard model.

In this analysis the mass of the Elko field was taken as $1.0$GeV and its effects during the inflationary phase is not important, thus the mass of the Elko field must be estimated by others observational constraints, as formation of structures for instance.  We have verified that an upper limit to the mass of $m\ll 10^{-10}m_{pl}\sim 10^9$GeV can be established if we want to keep the initial conditions before each epoch according to the standard model values. Such value is in agreement to modern estimates \cite{fuku}. It was also obtained that the number of e-foldings during the inflationary phase is strongly dependent on the initial value of the field, here taken as $\phi_i > m_{pl}$, as occurs in chaotic inflationary models. Inflation occurs exactly when the field goes from $\phi_i\sim m_{pl}$ to zero. As the field decays to the bottom of its potential it starts to oscillate and its amplitude diminishes with time.

As already pointed out previously, a similar analysis just for inflationary phase has been done in \cite{st}, using a phenomenological symmetry broken potential. The expected evolution as matter type after inflation has been obtained just by extrapolating the obtained curve and the effective mass estimated is strongly dependent on the parameters of the symmetry broken potential. Here both evolutions, namely radiation and matter type, were obtained using its expected evolution laws and initial conditions generated after inflation, showing a good agreement to standard model. Also, the mass used here for the field is free to be constrained by observations, since the inflationary phase is driven just by the quartic self-interaction term.

Although the exact analysis of the system of equations has been done only for a short time interval\footnote{The numerical analysis for the oscillatory behaviour of $\omega$ was verified from $10^{-35}$s up to $10^{-20}$s.}, the initial conditions generated after inflation for the scale factor and the energy density correctly lead to the expected values for the subsequent evolution. In this sense we can affirm that the Elko field drives all the evolution of the universe, including its action as dark matter in the last phase of evolution.

The Elko field considered here is just a classical field. As a final remark on the inflation driven by Elko field, maybe we could interpret the inflationary expansion as a consequence of a kind of Pauli exclusion principle or degeneracy pressure effect if the field is considered as a quantum one. When the fermionic quantum Elko field rolls down to the bottom of the potential, trying to occupy its minimal energy state, the degeneracy pressure acts expanding the whole system in order to separate the particles, once they can not occupy the same fundamental state. This could be the trigger for inflation occur. Such interpretation is not possible when inflaton field is of bosonic type. Also, after evolve as dark matter in late times, the field keep trying to occupy the minimum potential energy, but its quantum nature does not allows all particles with zero potential energy, thus a remaining net energy fills the whole universe, which could be interpreted as a kind of cosmological constant energy density, being responsible for the recent cosmic acceleration. Such possible quantum effect based on the Pauli exclusion principle for explain both inflation and recent cosmic acceleration driven by a mass dimension one fermion field deserves future investigations.

\begin{acknowledgements}
SHP is grateful to CNPq - Conselho Nacional de Desenvolvimento Cient\'ifico e Tecnol\'ogico, Brazilian research agency, for financial support, grants number 304297/2015-1 and 400924/2016-1. RFLH acknowledges financial support from CNPq (No. 303734/2014-0). We would like to thank professor J. M. Hoff da Silva for valuable discussions. 
\end{acknowledgements}


\end{document}